\documentstyle[preprint,prl,aps,psfig]{revtex}

\begin{document}

\def\ws{{$\omega_s$}}
\def\kvec{{\bf k}}

\bibliographystyle{prsty}


\title{Spin lifetimes of electrons injected into GaAs and GaN}

\author{Srinivasan Krishnamurthy}
\address{SRI International, Menlo Park, CA 94025}
\author{Mark van Schilfgaarde and Nathan Newman}
\address{Department of Chemical and Materials Engineering,
Arizona State University, Tempe, AZ}
\date{\today}
\maketitle




\begin{abstract}
The spin relaxation time of electrons in GaAs and GaN are determined
with a model that includes momentum scattering by phonons and ionized
impurities, and spin scattering by the Elliot-Yafet, D'yakonov-Perel,
and Bir-Aronov-Pikus mechanisms.  Accurate bands generated using a
long-range tight-binding Hamiltonian obtained from empirical
pseudopotentials are used.  The inferred temperature-dependence of
the spin relaxation lifetime agrees well with measured values in
GaAs. We further show that the spin lifetimes decrease rapidly with
injected electrons energy and reach a local maximum at the
longitudinal optical phonon energy.  Our calculation predicts that
electron spin lifetime in pure GaN is about 3 orders of magnitude
longer than in GaAs at all temperatures, primarily as a result of the
lower spin-orbit interaction and higher conduction band density of
states.


\vbox{\vskip 12pt}
\end{abstract}

Spin-based electronics utilizes the previously unexploited spin degree
of freedom in semiconductors for, potentially, improved non-volatile
memories, low power electronics, high speed analog and digital
systems, and quantum computing.$^{1}$ For these applications, the spin
polarized carriers can be created optically$^{2}$ in the semiconductor
or, preferably injected electrically$^{3-5}$ from a ferromagnetic
material.  One of the key aspects of this technology is that the spin
depolarization or relaxation lifetimes of these carriers is long in
the semiconductor for spin device operation.  These spin lifetimes can
be many orders of magnitude longer than the characteristic times
associated with conventional electronics, such as the transit time or
minority carrier lifetime and can facilitate improvements in the
device efficiency, sensitivity and/or speed. Despite the importance of
this parameter, there is only a limited amount of data available on
the spin lifetime of carriers in semiconductors.  Experimentally, the
spin lifetimes of low energy electrons in GaAs has been measured to
vary from 100 ns at 5K to 1 ns at 300K.$^{2}$ Spin relaxation
lifetimes are expected to be shorter for high kinetic energy electrons
and in defective material, but their exact energy dependence on these
parameters has not been established.  Since most traditional devices
emit electrons over potential barriers, resulting in hot carrier
injection into the device region, the information on energy-dependent
lifetimes is essential.  In addition, with a quantitative study, it
may be possible to offset the decrease in the lifetimes expected at
high energies by choosing a semiconductor material with smaller
spin-orbit (SO) coupling.

In this Letter, we report the results of our bandstructure-based
calculation of energy, temperature and doping-concentration dependent
spin lifetimes in GaAs and GaN. The calculated spin lifetimes of low
energy electrons are in reasonable agreement with measured
values.$^{2}$ We predict the spin lifetimes to decrease sharply with
electron energy. In addition, because of the large conduction band
density of states (DOS) and smaller SO interaction in pure GaN, the
spin lifetimes are predicted to be {\it three orders} of magnitude
longer than that in GaAs.

Spin relaxation by three mechanisms has been studied extensively in
the literature.$^{6-11}$ Elliot-Yafet (EY) scattering may be thought
of as the coupling between the two spin channels of a Bloch state by
the spin-orbit Hamiltonian.  D'yakonov-Perel (DP) scattering, present
in materials with spin-dependent conduction bands, is the interaction
involving the SO Hamiltonian and another perturbation, most notably
the phonon and impurity scattering. Finally, the Bir-Aronov-Pikus
(BAP) mechanism is a process involving the recombination of an
electron-hole pair; the photon is reabsorbed to create another
electron-hole pair of the opposite spin.  In our calculations, all
three mechanisms have been included. The SO interaction is calculated
from the well-known $k \cdot p$ formalism and treated as a
perturbation to the tight-binding Hamiltonian, along with other
spin-conserving scattering mechanisms such as impurity, longitudinal
optical (LO) phonon, and deformation potential scattering.  The spin
relaxation time could then be expressed in terms of the momentum
relaxation times.$^{6-12}$

We studied the spin relaxation of electrons using accurate GaAs and
zinc-blende GaN conduction bands generated with a long-range
tight-binding Hamiltonian obtained from empirical pseudopotentials.
We have successfully used this Hamiltonian to obtain bandstructures,
momentum and velocity scattering rates and studied high field
transport in many semiconductor compounds and alloys.$^{13,14}$
Excellent agreement was found between results obtained with these
scattering rates and those observed in GaAs hot electron
transistors.$^{15}$ Similar calculation of the momentum scattering
time $\tau_p$ is carried out for electrons injected into GaAs and GaN,
as a function of dopant density and electron temperature. They are
then used to calculate spin relaxation times limited by EY, DP, and
BAP mechanisms.$^{9}$ The spin scattering lifetimes by EY and BAP are
proportional to the momentum scattering lifetime.  Interestingly, in
the states with {\it shorter} momentum scattering times, the electrons
do not stay long enough for spin to precess, resulting in {\it longer}
relaxation times for spin scattering by DP mechanism. At very low
injection energies ($\le$ 10 meV), the leading spin scattering is by
EY (in n-doped) and BAP (in p-doped) material.  However the p-doping
must be at least 5 $\times$ 10$^{17}$ cm$^{-3}$ in GaN and 1 $\times$
10$^{18}$ cm$^{-3}$ in GaAs for spin scattering by BAP to be
comparable to that by EY. At higher energies, DP is the strongest
mechanism for spin scattering in both GaAs and GaN.  As the energy of
the electrons is increased, the DOS available for scattering also
increases.  However, the momentum scattering rate decreases as k$^2$
(for phonon scattering) and k$^4$ (for impurity scattering).
Consequently, the momentum scattering time increases with energy
initially, then falls steeply once the threshold for phonon emission
is reached. At even higher energy, the increase in DOS nearly
compensates the decrease in the matrix element, resulting in a
nearly-constant momentum relaxation time. The DP mechanism limited
spin relaxation time $\tau_s^{DP}$, which varies as
$\sim\tau_p$/E$_k^3$, decreases for increased energy but reaches a
local peak at the threshold energy for phonon emission. The product of
$\tau_s$ and the group velocity, ${\frac {1} {\hbar}} {\frac {\partial
E_k} {\partial k}}$, at that energy gives the energy-dependent mean
free path, $\lambda_s$ for the electrons.  The calculated spin mean
free path, $\lambda_s$, as a function of injection energy is plotted
for two donor doping concentrations and temperatures in Fig. 1 and
Fig. 2 for GaAs and GaN respectively. For all energies above 10 meV,
DP mechanism is dominant, and consequently, we see that spin
scattering lifetimes are longer at higher temperature and higher
doping density where momentum scattering lifetimes are shorter.  Also,
the mean free path decreases rapidly with energy and reaches a local
maximum at energy slightly above the LO phonon energy. Notice that
mean free path of electrons with one LO energy is about 0.1 cm in GaAs
and 1 cm in GaN. The longer mean free path of the injected electrons
in GaN is primarily attributed to the small SO splitting (12 meV in
GaN and 341 meV in GaAs), large band gap (of 3.4 eV in GaN and 1.5 eV
in GaAs), and large DOS (effective mass of 0.2 m$_0$ in GaN and 0.067
m$_0$ in GaAs). At higher energies, the mean free path continues to
decrease in spite of the increase in group velocity.

To determine the temperature dependence of the spin lifetime, a Fermi-
Dirac distribution can be safely assumed because the phonon-limited
momentum relaxation times ($\sim$ 50 ps at 5K in GaAs) is typically
two orders of magnitude shorter than the time scale of the spin
relaxation times ($\sim\mu$s).  An ionization energy for the donors in
GaAs of 5.8 meV is used, a value commensurate with the Si$_{Ga}$
defect.$^{17}$ As the lattice temperature is reduced, fewer impurities
are ionized, and the Fermi level (FL) is self-consistently
obtained. An photo generated carrier density typically used in
experiments,$^{2}$2 $\times$ 10$^{14}$ cm$^{-3}$, was added so that
these results can be directly compared to Ref. 2.  This last
assumption only affects the low temperature ($\le$ 10 K) result due to
carrier freeze-out.  Figure 3 shows the results obtained for GaAs and
GaN.  The calculated energy-dependent spin relaxation lifetimes can be
used to compare with the measured lifetimes at low or zero magnetic
field.  Figure 3 compares the averaged spin lifetimes (solid lines)
and the experimental values$^{2}$ (dashed-dots) in the temperature
range of 5 K to 145 K. We see that the predicted trend (and slope) is
in good agreement with experiment, except at very low temperature
($<$10 K). For T $>$10 K, the absolute value of the calculated values
are a factor of 2 to 3 smaller than experiment. At very low
temperatures, the predicted lifetime decreases with T, whereas the
measured lifetime appears to reach a plateau.  This difference may
result from the presence of compensating defects in the experimental
samples.  The dopant density and low field mobility of the GaAs,
reported in Ref. 2 at room temperature are 10$^{16}$ cm$^{-3}$ and
5400 cm$^{2}$/v.s, respectively.  Our calculations predict a mobility
of 7100 cm$^{2}$/v.s, in agreement with the values found for high
purity samples.$^{16}$ The lower mobilities measured in the experiment
indicate that defects are present in the samples. The defect states
would be expected to influence the FL and scattering times,
particularly at low temperatures.

Assuming 5.8 meV of donor ionization energy for Si in zinc-blende GaN,
the averaged spin lifetime calculations are carried out, and the
values are also shown in Figure 3 (thick solid line). It is
interesting to note the spin lifetimes in GaN are expected to be about
three orders of magnitude larger than those in GaAs. The predicted
enhancement is because of the combination of the dominance by DP
mechanism, larger momentum scattering and smaller SO interaction.
However, when the defect density is large, the defect-assisted EY
mechanism which is dominant at low temperatures, severely limits the
spin lifetimes. This is in agreement with the recent results in highly
impure (with 5 $\times$ 10$^{8}$ cm$^{-2}$) wurtzite-GaN where the
measured lifetimes are an order of magnitude smaller than that in GaAs
at T$\le$ 50K.$^{17}$ When the material quality is improved$^{18-20}$
for much lower defect density ($<10^{6}cm^{2}$), a large increase
in spin lifetimes is expected.

In summary, we have calculated the energy and temperature dependent
spin lifetimes of electrons using full band structures of GaAs and
GaN. The calculated spin lifetimes of low kinetic-energy electrons in
GaAs are in reasonable agreement with measured values.$^{2}$ We
predict the spin lifetimes to decrease sharply with electron energy
and reach a local maximum near one LO phonon energy at low T. In
addition, because of the large conduction band DOS and smaller SO in
GaN, the spin lifetimes are predicted to be {\it three orders} of
magnitude longer than that in GaAs. When non-radiative recombination is
not dominant, the mean free path for the electrons with an injected
energy of 0.5 eV is predicted to be about 1 mm in high-purity GaN,
suggesting GaN as more suitable materials for spin injection devices.
The accuracy of the model will be enhanced by using parameter-free
calculations in which the SO coupling is included in the Hamiltonian,
and all other scattering processes (electron-phonon,
electron-impurity) are considered as perturbations.  Such calculations
are in progress.

One of the authors (S.K) thanks Marcy Berding, Arden Sher, and Michael
Flatte for useful discussions and Jay Kikkawa for providing the spin
lifetimes data. This work was supported by the DARPA. (ONR contract \#
N00014-02-1-0598).

\newpage

\begin{figure}
\mbox{\psfig{file=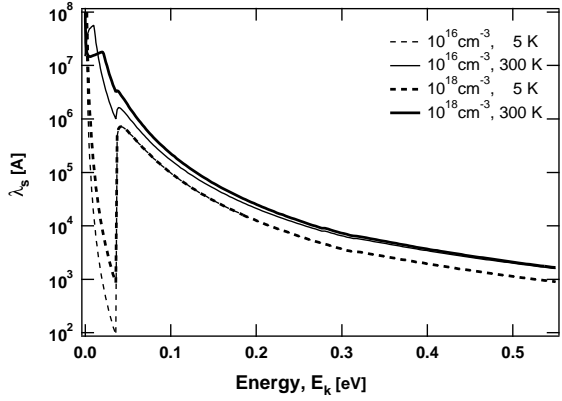}}
\caption{Calculated spin mean free path in GaAs.} 
\end{figure}

\newpage
\begin{figure}
\mbox{\psfig{file=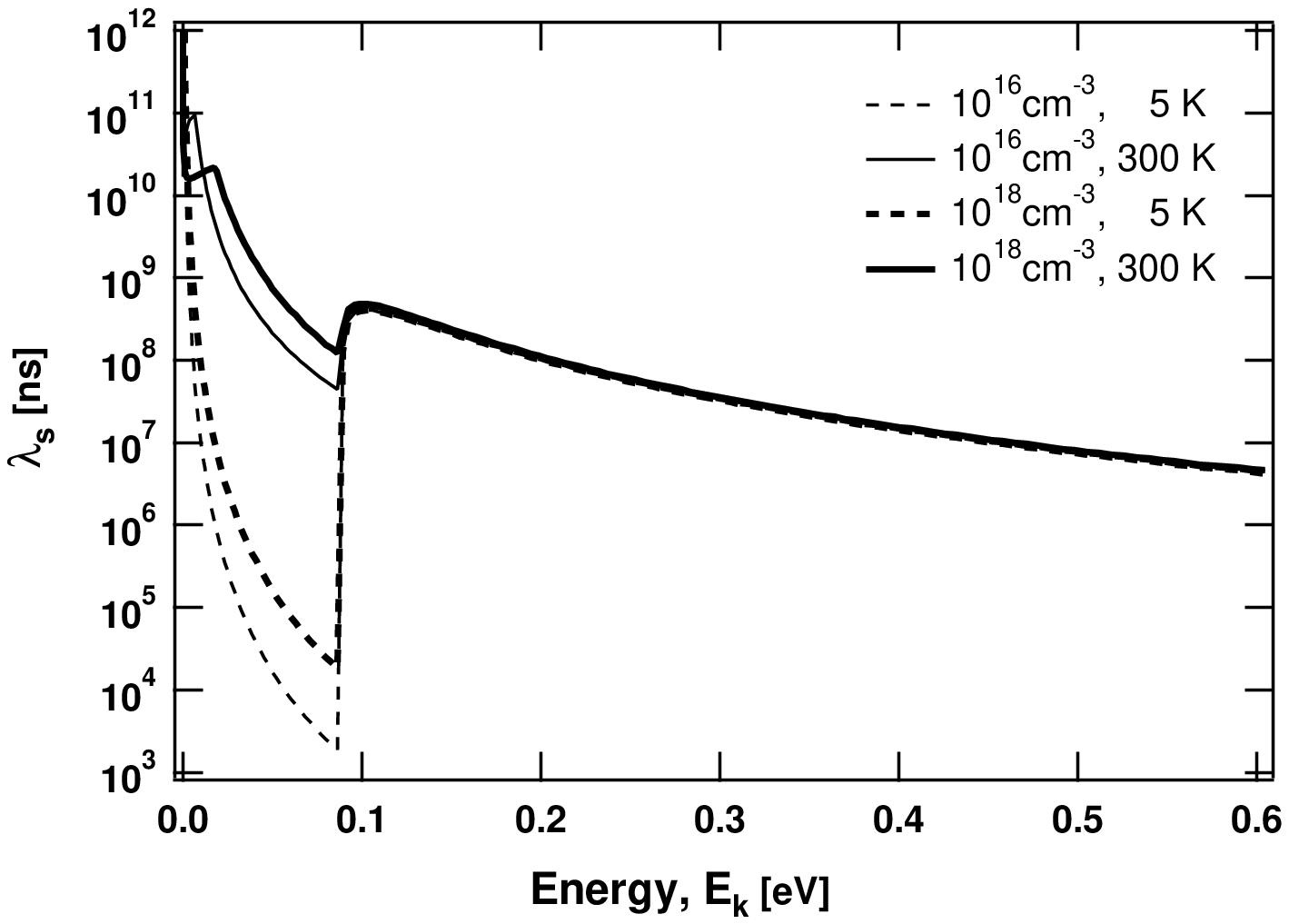}}
\caption{Calculated spin mean free path in GaN.} 
\end{figure}

\newpage
\begin{figure}
\mbox{\psfig{file=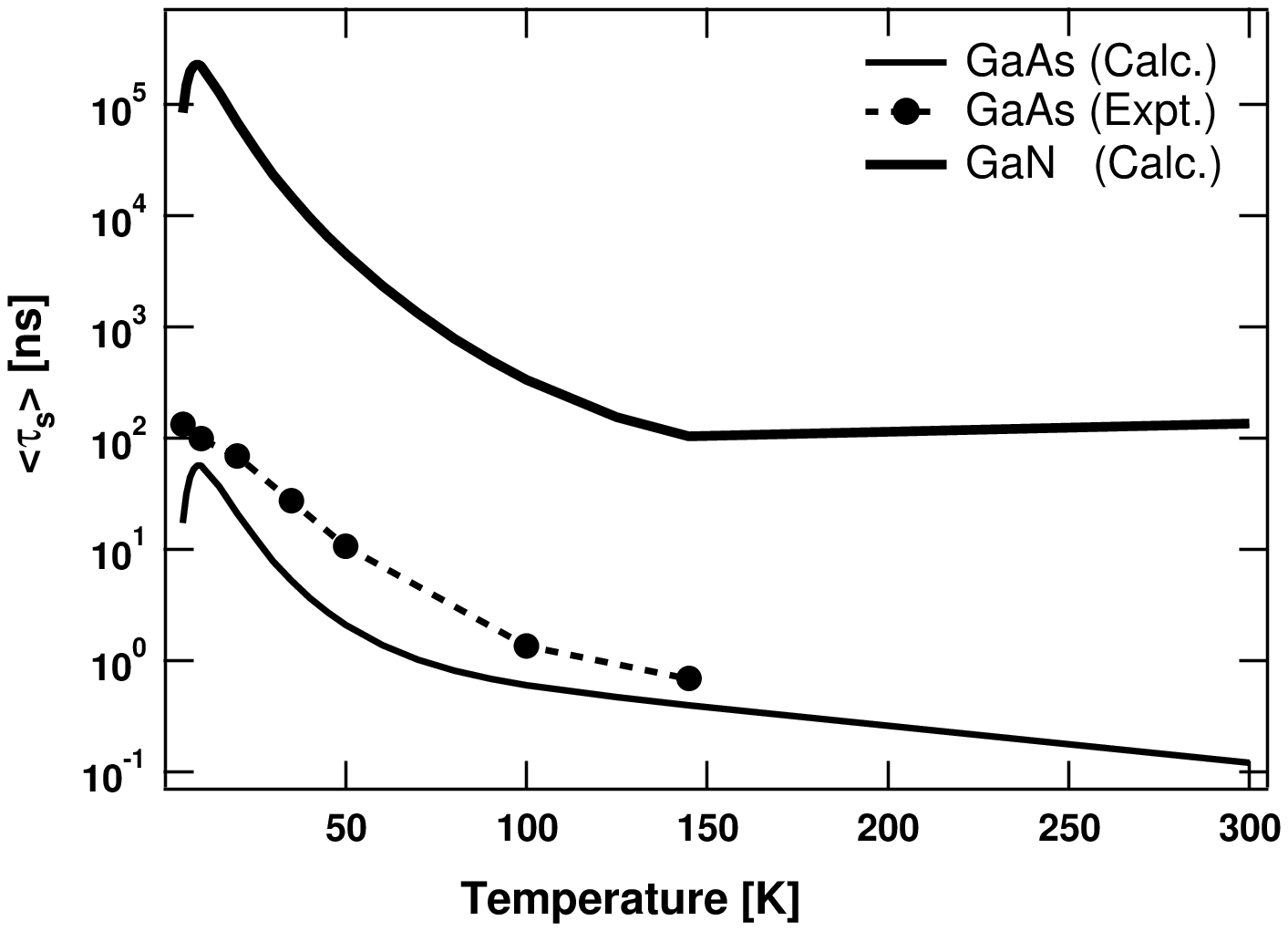}}
\caption{Spin lifetimes vs temperature in GaAs and GaN.}
\end{figure}

\end{document}